\documentstyle[aps,twocolumn]{revtex}
\input{epsf}

\def\be{\begin{eqnarray}}
\def\ee{\end{eqnarray}}

\def\d{\delta}

\def\G{\Gamma}
\def\Gt{\tilde G}
\def\l{\lambda}
\def\o{\omega}
\def\O{\Omega}
\def\cg{c^{\dagger}}
\def\dg{\dagger}
\def\kt{\tilde k}
\def\s{\sigma}
\def\e{\epsilon}
\begin{document}
%
%
%
%
\title{ Fragmentation of bands and Fermi surfaces in stripe phase.}
\author{ M.Ya.~Ovchinnikova.}
\address{{\it Institute of Chemical Physics, RAS,
Moscow, 119977,  Kosygin str., 4.}}
\wideabs{
\maketitle
%
\begin{abstract}

The mean-field study of the stripe phases of $t-t'-U$ Hubbard model confirms
formation of the in-gap "impurity" subbands of states localized on domain
walls.  For bond-aligned stripes it is shown that segments of the Fermi
surfaces in antinodal and nodal directions correspond to the "impurity"
band or that delocalized over a whole antiferromagnet domain. This might
explain the dichotomy between corresponding quasiparticles in $La_{2-x}Sr_x
CuO_4$. The Fermi surface changes its topology
 at some doping depending
on stripe width. Additional ordering at the domain wall should be
supposed to accord the observed $1\over 8$ anomaly and the stripe spacing. It
is confirmed also that the diagonal stripes can provide the dielectric state
at nonzero doping.

\end{abstract}
}


The angular resolved photoemission (ARPES) makes a direct probe of the
quasiparticles (QP) in  cuprate superconductors \cite{1}. Thus, the d-wave
superconducting (SC) gap and the pseudogap (PG) in normal state was
measured in underdoped (UD) Bi-based cuprate (BSCCO). The PG opens in
antinodal directions similarly to SC gap. There are many arguments
\cite{1} which connect the PG origin with properties of the correlated
antiferromagnet (AF) Fermi liquid, namely, with a profile of the lower
Hubbard band. This profile is well traced in undoped dielectric compounds.
At small h-doping this profile determines a hole pockets centered at
$S(\pi/2,\pi/2)$ and the PG in excitation spectrum at $k$ near points
$M(\pi,0)$. The homogeneous solutions of $t-t'-U$ or $t-t'-J$ models confirm
such picture and predict a change of the FS topology from small hole pockets
to large e-like FS.

However the observed FS evolution with doping contradicts the e-like FS.
For $La_{2-x}Sr_x CuO_4$ (LSCO) the FS arc retains a hole-like character and
the FS always crosses the line $M(0,\pi)-Y(\pi,\pi)$. The ARPES study of LSCO
shows \cite{2,3,4,5,6} that there are two different types of the FS
segments.  The first ones are the straight segments parallel to CuO bonds
near points M.  The second ones are the segments of arc near the diagonal
nodal directions with a systematically suppressed photoemission intensity.
The dichotomy between the nodal and antinodal quasiparticles have been
revealed in the UD LSCO \cite{4,5}.  An extra broadening of quasiparticles
was observed starting from points $k=(k_0,\pi-k_0)$, $k_0=0.175\pi$ where the
FS turns from diagonal direction $M(\pi,0)-M'(0,\pi)$ to $\G-M$ directions.

In \cite{4,5} the FS behavior in LSCO was connected with the stripe phase -
the structure of antiphase AF domains separated by domain walls and aligned
in one direction. The experimental evidence of such quasistatic structure
comes from incommensurate (IC) peaks in inelastic neutron scattering. For
LSCO the IC peaks have been observed at $Q=(\pi,\pi)\pm(\d,\pm\d)$ or at
$Q=(\pi\pm\d,\pi),~~(\pi,\pi\pm\d)$  in dielectric or superconducting phases
correspondingly \cite{7,8,9}.  Comparative study of the IC peaks in
elastic and inelastic scattering \cite{10} supports the existence of slowly
fluctuating stripe structures.  Some features of ARPES spectra have also been
explained by the stripes.

Another type of the PG anisotropy have been predicted \cite{11,12} and
observed \cite{13} in ARPES data for e-doped cuprates $Nd_{1-x}Ce_xCuO_4$
(NCCO) in UD region ($x<0.1$). The e-pocket around points M arise and the PG
opens at diagonal direction in accordance with the energy profile of the
upper Hubbard band along the boundary of the magnetic Brilluin zone. But at
higher doping the FS manifests fragmentation onto separate segments near
$M(\pi,0)$ and segments at diagonal directions \cite{14}. Such
patch-like character of the FS was ascribed to coexisting e- and h-pockets in
upper and lower Hubbard subbands \cite{15} or to the stipes \cite{16}.

The aim of present work is to show that the fragmentation of the bands and
the FS's may be caused by stripe formation. For periodic stripe
structure the FS fragmentation is due to splitting of original band into the
several subband, unlike a splitting into the lower and upper
subbands only in homogeneous AF state. The  quasi-onedimensional (1D)
"impurity" bands are placed inside the gap between lower and Hubbard
subbands. The transfer of the spectral weight between bands are calculated
and discussed in light of ARPES data for LSCO. Full population of
"impurity" subband of periodic structure leads to the
its anomalous properties at some doping similar to ${1\over 8}$-anomaly in
LSCO.

Influence of the stripes on a shape of FS and spectral intensity map of
photoemission have been studied earlier in \cite{17,18,19}. There the
stripes was described by a hand picked spin potential. In this way the
bond-aligned or diagonal stripes with the bond- or site-centered domain walls
were considered. One of the main feature obtained was an appearance of the
in-gap states. These states according \cite{17,18,19} are characterized by
localization of charge near the domain walls and quasi-one-dimensional (1D)
type of dispersion over momentum component parallel to the stripe.

To check a stripe phase stability and to trace an evolution of spectral
features with doping we use the mean-field (MF) approach
instead of the hand-picked stripe potential. Here we present a brief report
of results, more details will be given elswhere. Consideration is
limited by periodic structures without any disorder.  Earlier \cite{20} it
was shown that in MF approximation a homogeneous AF state of the 2D $t-t'-U$
Hubbard model is unstable relative to the stripe formation in case of the
h-doping, but it remains the most stable state for the e-doped models.

Consider a t-t'-U Hubbard model with the zero band energies
$\e_k$. The MF approximation is insufficient to describe the superconducting
(SC) pairing.  But for simplest study of the normal state structures we
retain the original $t-t'-U$ model.  A periodic structure is determined by
twodimensional translation vectors $E_i$ and $B_i$ in direct and inverse
lattices which satisfy the relations ${\sl E}_i B_j=2\pi \d_{ij}$. The
components of ${\sl E}_i$ ¨ $B_i$ are in units of the corresponding lattice
constant $a$.  The unit cell of the structure contains $n_c$ sites and the
site positions  $n=n(L,j)$ is determined by the integers $L=(L_1 , L_2)$ and
$j=(j_x,j_y)$ which determine the positions of the unit cell and of a site
inside the unit cell.

Any quasimomentum $k\in G$ inside a full Brilluin zone (BZ) of original
lattice is expressed via momentum $\kt$ reduced to the main BZ $\Gt$ of the
periodic structure: $k=\kt+B_1m_1+B_2m_2$. A set of integers ${m_1,m_2}$
numerate all independent Umklapp vectors. The area
of $\Gt$ are restricted by conditions $|\kt B_i|\leq B_i^2/2$.
The order parameters (OP) of the periodic structures are
the mean charges and spins on sites $j$ of the unit cell
\be
r_j=<r_{n(L,j)}>;
~~S_{z, j}=<S_{z,n(L,j)}>
\label{1}
\ee
The MF solutions for periodic structure is obtained in standard way
\cite{20}.  The mean energy  $<H>$  is an explicit function of OP
(1). A wave function is determined by occupation of one-electron eigenstates
$\chi_{k\l}^{\dg}$ of the linearized  Hamiltonian
$H_{lin}=\sum_{{\kt}\in\Gt}{\hat h}_{\kt}$.  The latter is divided into
independent contributions from each reduced quasi-momentum $\kt$.  In the
momentum represen\-ta\-tion the eigenstates are expanded
over a basis set of $2n_c$ Fermi operators
\be
\chi_{{\kt}\l}^{\dg}=\sum_{m,\s}\cg_{\kt+Bm,\s}W_{m\s,\l}(\kt ),
\label{2}
\ee
where $\l=1,\dots,2n_c$, $Bm=B_1 m_1+B_2 m_2$  and vectors ${\kt+Bm}$
share all phase space $G$.

The matrix of eigenvectors $W_{m\s,\l}$ and  the eigenvalues $E_{\kt,\l}$
are determined by diagona\-li\-za\-tion of $h_{\kt}$  in the basis set
$\{\cg_{\kt+Bm,\s}\}$.
The order parameters (1) are in turn expressed via the matrix
$W^*_{ms,\l}(\kt)$ and the Fermi functions $f(E_{\kt \l}-\mu)$.

The intensity of the photoemission of electron with a projection $k$
of the momentum on $ab$ plane and the energy $\o=E-\mu$ is
\be
I(k,\o)=|M(k)|^2 A(k,\o)f(\o)\otimes R_{\o k}.
\label{3}
\ee
It is determined by the matrix element $M(k)$,
a spectral density $A(k,\o)$ and the Fermi function $f$.
Usual convolution is done with the Gaussian function
$R_{\o k }$ \cite{21} with parameters, which characterize  a finite
resolution over $k$ and energy.
Here we take a constant value for matrix element $M$
since we study an effect of the structure on the spectral density $A$.
In the one-electron approximation
\be
\begin{array}{ll}
A(k,\o)=&{1 \over N} \sum_{\kt\in\Gt}\sum_{m,\s,\l}
|W_{m\s,\l}(\kt)|^2\\
&\times {\overline \d} (E_{\kt \l}-\mu-\o) \d_{k,\kt+Bm}
\end{array}
\label{4}
\ee
Here $\l=1,\dots , 2n_c$ and index $m=(m_1 , m_2 )$ numerates all
independent Umklapp vectors $Bm$.
A standard replacement of the $\d$- function in Eq.(9) by a
function with finite
width $\O$ is implied. A map of $I(k_x,k_y, \o=0)$ visualizes both
the main and shadow segments of the FS. Though the band energies are
periodic functions in $k$ space, the intensity (3) has not such
periodicity.  Therefore various segments of the FS display themselves with
different intensity  even for the constant matrix element $M$ in Eq.(3).

Calculations were carried out for vertical stripes $Y_l$ with the
bond-centered domain walls and spacings between them $l=4,6,8,10$ (in lattice
constant).  The structures correspond to numbers $n_c =2l$ of sites
in unit cell.  The model parameters are $U=4$, $t'=0.1$ (in unit $t$).

Fig.~1a presents the mean energies per one site as functions of doping for
the vertical stripe phases in compare with energy for uniform AF state.
For h-doped models (unlike e-doped ones) the stripe phases are more stable
than AF state \cite{20}. With doping the lowest
structures change each other in the sequence $Y_10, Y_8, Y_4$, but not $Y_6$.
In case of the stripe spacing $l=8,10$ the doping dependencies ${\overline
H}(p)$ have a break of slope at some doping $p'(l)$ at which the additional
two domain walls appear inside the stripe width. So the width of truly AF
domain narrows to $l=4$ or $5$. But at doping $p'$ the structure $Y_4$
is lower in energy than $Y_{10},Y_8$. Note that the MF treatments gives only
approximate sequence of most stable stripe phases. Nevertheless it
demonstrates the preference of structure $Y_4$ in wide range of doping.

The maps of intensities $I(k_x,k_y, \o=0)$ in Fig.~2 present the Fermi
boundaries for structure $Y_8$. They are typical for all bond-aligned
stripes. At doping $p\le {1\over l}={1\over 8}$ the FS consists of quasi-1D
segments near points $M(\pm\pi,0)$ and these segments are normal to the
stripe direction.  Fig.2b presents the FS symmetrized over $x$- and
$y$-aligned stripes.  Absence of the FS at the diagonal cuts means an opening
of the PG at the nodal direction unlike a standard PG at antinodal regions in
BSCCO. For structure $Y_4$  the antinodal parallel FS segments display itself
in large doping range. Such behavior is consistent with the
FS observed in ARPES spectra of the UD LSCO \cite{4,5}.

For the phase $Y_l$ at large doping $p>{1\over l}$ the FS changes its
topology: the antinodal segments of FS disappear and the nodal FS segments
arise around the hole pockets as it is shown in Fig.~2c.

Origin of such behavior becomes clear from an
analysis of the electronic bands of stripe phases \cite{17}.
The antinodal quasi-1D segments of FS has been acquired to the
band appearing inside the Hubbard gap. This is an "impurity" band of states
localized at the domain walls.  In \cite{16,17} the
stripes have been modeled by a hand picked spin potential.  For vertical
stripes  this in-gap band have large $\sim t$ dispersion along y axis. The MF
calculations of stripe phases confirm appearance of the in-gap "impurity"
band with large dispersion.

Figure 3 presents a map of spectral function $A(k,\o)$, Eq.(2), for structure
$Y_8$ at critical doping $p={1\over 8}$. The bands of periodic structure
and a transfer of the spectral weight between them are shown for k varying
along the contour $\G-M(\pi,0)-Y(\pi,\pi)$. It is seen that a level of
chemical potential falls just into the small gap between the "impurity" band
which is fully empty at $p={1\over 8}$ and the other seven lower levels of
"main" bands which are completely occupied. Corresponding states are
delocalized over all sites of the AF domain.
At doping $p<{1\over 8}$ the antinodal FS segments originate from "impurity"
band, whereas at $p>{1\over 8}$ the FS segments at nodal direction are due to
"main" bands from whole AF domain.

More realistic description of LSCO imply an ensemble of quasistatic
structures with stripes of different width. One should also take into account
the fluctuations of charge density over a space and times. Some simplified
models of disordered stripes has been tested in \cite{20}. The charge
fluctuations or phase separation onto the poor and rich doping regions and
variations of stripe structures might lead to situation when  both the
nodal segments and the antinodal quasi-1D segments are seen in the intensity
map simulteneously.

The above analysis allows to state that different types of the FS
segments have different origin. We may suppose that the antinodal segments of
quasi-1D band are more influenced by the structure defects and impurities
than the nodal FS segments corresponding to bands delocalized over the AF
domain. This may explain a difference in the shape of the ARPES signals
obtained for these segments in LSCO \cite{6,7}. In energy distribution
curves the sharp peaks have been observed or not observed in the nodal and
antinodal regions of $k$ correspondingly. From the above FS fragmentation it
follows that near the crossing points of the standard FS arc with the MBZ at
$k\sim (\pi-k_0,k_0)$, $k)-\sim 0.17\pi$, there is the FS regions with a
suppressed photoemission intensity and with a broadened QP peak. In
\cite{5} such points are called the "hot" points, since the broadening of
the QP peak starting from these points are acquired to setup of additional
mechanism of the QP scattering. In concept of stripe phase, the disappearance
of sharp FS at hot points are due to a transfer of spectral weight between
the FS segments of different nature - from band states delocalized over AF
domain to 1D "impurity" band which is more sensitive to the structure defects.

It is attractive also to search a connection between the anomaly of
electronic properties of the stripe structure $Y_8$ at $p={1\over 8}$ and the
observed anomalous suppression of superconductivity and $T_c$ in LSCO at such
doping.  If we suppose that the structure $Y_8$ gives the main contribution,
then suppression of $T_c$ might be explained by opening of the PG along the
whole expected FS at such doping in a situation when the in-gap "impurity"
band is entirely free and the other 7 lower delocalized bands are entirely
occupied.  However the argument against the
structure $Y_8$ is that the IC peaks in inelastic neutron scattering are
observed at momentum $q=\pi(1\pm{1\over 4},1)$ in difference from
$q=\pi(1\pm{1\over 8},1)$ expected for the structure $Y_8$ with stripe
width $l=8$. As to simple structure $Y_4$, one cannot expect any anomaly in
its conductivity since the "impurity" band is only half-filled for
the structure $Y_4$.  The idea that the observed ${1\over 8}$-anomaly is
connected with complete occupation of 7 bands among 8 possible bands together
with the observed stripe width $l=4$   require an assumption about aditional
ordering the domain wall states.  It might be an alternating of valence bonds
along domain wall which split the "impurity"  band into two  in-gap bands
localized on the domain walls. A search and study of suitable models
with an ordered system of valence bonds in addition to the spin stripes are
needed in order both to agree the ${1\over 8}$-anomaly of $T_c$ with a stripe
period and to establish the mechanism and symmetry of the SC pairing in LSCO.

It is instructive also to stress a difference between diagonal and
bond-aligned stripes. Existence of diagonal structures in dielectric phases
of LSCO is evidenced by peaks in INS at $Q=(\pi\pm\d,\pi\pm\d)$ at
small doping $p<0.05$ \cite{9}. We calculated the structures $D_l$,
$l=8,10,20$, with diagonal stripes and with spacing $l/\sqrt{2}$ between the
domain walls.  We choose stripes with the site-centered domain walls, as
structures with lower energy.  Corresponding translation vectors and a
number of sites in unit cell are $E_{1,2}=(-1,1),(l,l)$ and $n_c=2l$.

Fig.4a presents the FS segments on the map of intensity $I(k,\o=0)$ for
structure $D_8$ at doping $p=0.15>{1\over l}$. In accordance with predictions
\cite{20} the FS segments are normal to the stripes. Figures 4b,c present
the band dispersions for $k$ varying along both diagonals as they display in
the map of spectral functions. The "impurity" band associated with states
localized at the domain walls lies again inside the Hubbard gap.  Unlike
similar band in vertical stripes, for diagonal stripes this "impurity" band
has a small dispersion ($\sim t'$) and is separated from lower bands by
significant gap. This means that at doping $p={1\over l}$ the diagonal
stripe phase $D_l$ would be a dielectric one since the entirely occupied
lower bands are gaped from the empty "impurity" band. Such arguments first
proposed in \cite{18} explain how a dielectric phase may exist even at
nonzero doping. Really the dielectric phase of LSCO is characterized by large
mean spacing of domain walls $l>20$ \cite{9}.

The question on a position of the chemical potential $\mu$ of the doped
dielectic phase inside the Hubbard gap is actual in connection with new
ARPES study of $Ca_{2-x}Na_xCuO_2Cl_2$. It may be supposed that its
position coinsides with a position of the "impurity" band of states localized
on the diagonal domain walls. Then the energy $E-\mu<0$ of the wide QP hamp
would be determined by a gap between "impurity" band and the lower
Hubbard subbands. Then the width of the QP hump may reflect a degree of a
stripe disorder.  It would be interesting to search the diagonal stripes
in this compound. A more sharp edge of the QP hamp at $p=0.1$ might imply  a
unification of stripe structures to stripe phase with definite spacing.

In conclusion, the stripe formation leads to fragmentation of bands and the
Fermi surfaces, namely, to separation of "impurity" band into the Hubbard
gap. This quasi-1D band corresponds to states localized on the domain walls.
For vertical stripes this band gives origin of antinodal segments of FS,
whereas the nodal segments of FS are due to the rest lower subbands
corresponding to states delocalized over the whole AF domain. Different
nature of these FS segments might explain the dual charachter FS segments in
UD LSCO, suppression of the nodal FS segments and dichotomy between the nodal
and antinodal quasiparticles as observed in ARPES.

 The stripe solutions with spacing $l$ change the FS topology
at doping $p=1/l$ at which the isotropic pseudogap is opened.  In
light of this effect the $1\over 8$ anomaly and the observed stripe spacing
in LSCO imply an additional ordering (may be bond alternating) along the
domain wall. For diagonal stripes the small dispersion of the in-gap
"impurity" band can provide a dielectric phase of doped materials.

Work is supported by the Russian Foundation of Basic Research under Grant
No.03-03-32141. Author is grateful to V.Ya.Krivnov for stimulating
discussions.

%

\newpage

Captions to figures.

Fig.1.

a - the doping dependencies of the mean energy (per one site)
of the hole doped model with parameters $U=4.0$, $t'=0.1$
for homogeneous AF solution and for the vertical stripe phases
$Y_l$.  The corresponding curves are marked by values $l=4,6,8,10$. b - the
same at small doping for diagonal stripes $D_{10}, D_{20}$; dashed
segment correspond to incomplete convergence of MF procedure. c- variations
charge and spin densities (1) as functions of site number $n_x$ for the
structures $Y_8$ and $D_8$ (solid and dashed curves).

Fig.2.

a - the map of the photoemission intensity $I(k,\o=0)$, Eq.(3), displaying
the FS for the stripe structure $Y_8$ at the doping $p=0.1<\frac 1 8$. b -
the same symmetrized over the x- and y- orientations of
stripes. c - the FS for the same structure at $p=0.15>\frac 1 8$.
Critical doping $\frac 1 8$ corresponds to opening small PG along
the whole expected FS.

Fig.3.

Behavior of bands  $E_\l(k)-\mu$ for structure $Y_8$  at critical doping
$p={1\over 8}$ for $k$ varying along the contour $\G-M(\pi,0)-Y(\pi,\pi)$
on the map of the spectral density (bottom) or nonweighted density of state
(top). Horizontal line marks the level of chemical potential.

Fig.4.

a - the Fermi surfaces for diagonal stripe structure $D_8$ on the map of
intensity $I(k,\o)=0$ at doping $p=0.15$. Only the FS segments normal
to stripes manifest.  b,c - dispersion of bands for $k$ varying along
diagonals $\G -Y'(-\pi,\pi)$ and $\G-Y(\pi,\pi)$ as seen on the map of the
spectral function for same structure.

\end{document}